\documentclass[aps,prb,floats,twocolumn,showpacs]{revtex4}
\usepackage{latexsym}
\usepackage{graphicx}

\usepackage{epsfig}
\usepackage{amsmath}
\usepackage{color}
\newcommand{\red}{\textcolor{red}}

\begin{document}

\title{Magnon excitations in $Cs_2CuAl_4O_8$ - a bond alternating S=1/2 spin chain with next nearest neighbor coupling} 
\author{Satyaki Kar}
\affiliation{Department of Theoretical Physics,  Indian Association for the Cultivation of Science, Kolkata 700032, India.}  
\date{\today}
\begin{abstract}
A recent density functional theory (DFT) based analysis, complemented with Quantum Monte Carlo calculations\cite{main} revealed a highly spin-frustrating nature of the one-dimensional spin-$\frac{1}{2}$ compound $Cs_2CuAl_4O_8$ that comprises of unique bond alternations and relatively strong next nearest neighbor interactions. This article gives a brief account on possible magnon excitations that can appear in the ground state of such systems. We find that the spin waves obtained on top of coplanar helical reference states show multiple magnon modes (both acoustic and optical). However, those magnon modes turn out to be stable only in the absence of bond alternations.

\end{abstract}
\maketitle

\section{Introduction}

Recently synthesized\cite{ref12} cesium-copper-aluminate compound $Cs_2CuAl_4O_8$, containing a zeolite-like structure with separated edge-sharing chains of copper ions, has already been probed thoroughly using first principle calculations\cite{main}. This points out to an highly frustrating spin chain arrangement within the crystal.
The spin model for $Cs_2CuAl_4O_8$, as obtained from first principle calculations  and backed up with quantum Monte Carlo results\cite{main}, reveals a one
dimensional S= 1/2 chain with unique bond alternations (observed in every third bond, as seen in Fig. 1) as well as strong next nearest neighbor spin interactions given as,
\begin{align}
H&=J\sum_i^{N/3}[S_{3i-1}.S_{3i-2}+S_{3i}.S_{3i-1}]+J^{'}\sum_i^{N/3}S_{3i}.S_{3i+1}\nonumber\\
&+J_{nnn}\sum_i^{N/3}S_{3i}.S_{3i-2}+J_{nnn}^{'}\sum_i^{N/3}[S_{3i+1}.S_{3i-1}+S_{3i}.S_{3i+2}]
\label{eq1}
\end{align}
where $i$ represents the position of Cu site in the chain.
In fact, Ref.[2] considers $J'=-J=J_{nnn}/3=26~K$ in such a model to mimic the low temperature susceptibility of the $Cs_2CuAl_4O_8$ crystal. 
In the following we give a brief analysis using  spin wave theory (SWT) in order to understand the
elimenatry excitations in such systems.

%\section{Exact Diagonalization results}
%We may exact diagonalize the model Hamiltonian to check for the behavior of the lowest energy modes as well as the lowest energy gaps for a range of parameters. Using LAPACK, an exact diagonalization (ED) is performed on the system Hamiltonian (with $J'_{nnn}=J_{nnn}$) for chain length of $L=12$.
%Results are shown in Fig.\ref{ED1} and Fig.\ref{ED2}. 
%In Fig.\ref{ED1}, the gap ( per unit length) between ground state and the lowest excitation are shown for $J_{nnn}=3$. We see that
%the values obtained are more-or-less constant with $\Delta\sim 0.16-0.17$, which is larger than the computational error $\sim 1/L=0.083$ due to finite sizes.
%So a small (but significant) gap is registered from ED calculations.
%We find this gap to decreases further as the strength of $J_{nnn}$ is reduced more and more.
%In the parameter set compatible for the compound Cs$_2$CuAl$_4$O$_8$, $i.e.,$ at $J/J'=-1$ and $J_{nnn}/J'=3$, our
%calculation gives $\Delta/J_{nnn}=0.055$, which is rather close to the QMC estimate of $\Delta/J_{nnn}=0.043$.
%Furthermore, we find that as $J_{nnn}$ is reduced the ground state energy $E_g$ gets lower in proportion (see Fig.\ref{ED2}).

\section{Spinwave Excitations}
In one dimension, Mermin-Wagner theorem forbids breaking of any continuous symmetry and thus there can not be any spontaneously 
broken symmetric phase possible. For the same reason, no magnetic-sublattice structures are possible either.
A continuum of S=1 elementary excitation spectrum is observed which results from spinon excitations in pairs, with relative momentum between 
them as 
an internal degrees of freedom\cite{pearson}. This can also be described as particle-hole excitations in a Jordon-Wigner transformed fermionized picture\cite{fad}.
This continuum lies between a lower boundary $\epsilon(k)=\frac{\pi}{2}J|sin k|$ and a higher boundary of $\epsilon_2(k)=\pi J|sin(k/2)|$
in an isotropic Heisenberg antiferromagnet.
The lower boundary $\epsilon(k)$ also represents the lowest magnon excitation spectra as obtained by des Cloizeaux and Pearson\cite{pearson} 
(dCP).
\begin{figure}
\includegraphics[width=.98\linewidth]{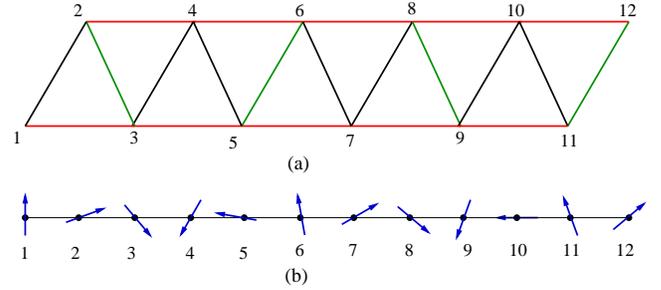}
\caption{Cartoons showing (a) various bonds in the $J-J'-J_{nnn}$ chain and (b) a typical helical spin reference state. Black, green and red colored bonds
 represent $J,~J'$ and $J_{nnn}$ couplings respectively whereas the spins are denoted by blue lines with arrowhead.}
\label{figstg}
\end{figure}

Surprisingly, Anderson's magnon excitation modes are only a $\pi/2$ factor away from this famous dCP expression, although in Anderson's picture,
antiferromagnetic sublattice structure is indeed presumed.
(we may mention here that fermionizing the spin Hamiltonian of the chain using a Jordon-Wigner transformation (JWT) gives better estimates for elimentary excitations. An Heisenberg NN spin interating chain gives $\epsilon(k)=(1+\frac{2}{\pi})$ cos$k~$\cite{breinig}, which is a better estimate as compared to the Anderson's spin wave approach and matches closely with famous dCP expression).
This similarity between magnon excitation spectra from Anderson's spin wave approach and that by a 
more rigorous method by Bethe and Hulthen\cite{pearson}, in the thermodynamic limit, makes the former an simple yet useful tool 
to understand the magnetic excitations in a spin-chain.

Apart from that, in real 3D materials its very rare to realize pure 1D spin chains, as interactions from transverse directions, even if very 
weak, contributes causing a quasi-dimensionality to the problem and hence a magnetic ordering can be presumed for the calculations. 

To begin with the Heisenberg spin chain in the Anderson's spin wave theory, 
we first note that the magnon dispersion is $\omega_k=J|sin(k)|$ about the Neel ordered 
reference state, containing two degenerate acoustic branches within the reduced Brillouin zone (RBZ) $(-\pi/2,\pi/2)$.
Now let us switch on the next nearest neighbor interaction $J_{nnn}$. The resulting $J-J_{nnn}$ model becomes spin-frustrated for an antiferromagnetic (AF) $J_{nnn}$.
It can be shown that for large $J_{nnn}$, no
stable doubly staggered spin reference states ($i.e.,~\uparrow~\uparrow~\downarrow~\downarrow~...$) are obtained, unlike
columnar reference states observed within the same limit in two-dimensions\cite{kar}.  The phase-point $J_{nnn}=J_{nn}/2$
is called the Majumdar-Ghosh point\cite{ref11} where the ground state obtained is  a dimer product state and the entanglement of the pair of next-nearest
neighbor spins shows a maximum\cite{ravindra}.
\begin{figure}
\includegraphics[width=.9\linewidth,angle=0]{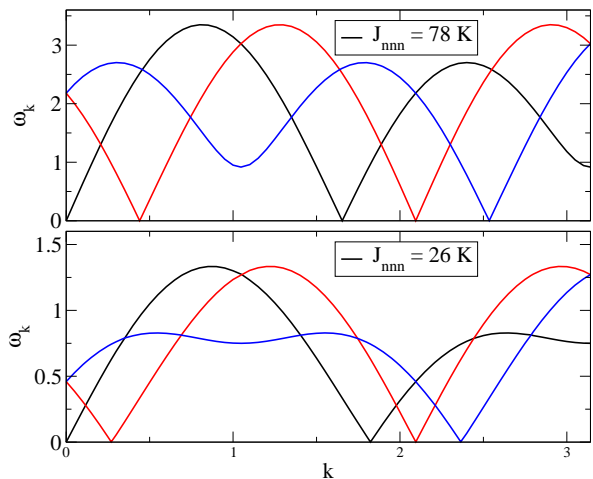}
\caption{Magnon modes about the helical reference state for $J = J' = 26~K$ (in units of $k_B$).
$\omega_k$'s are shown in units of $J$.}
\label{mag-modes}
\end{figure}

Given such background, we understand that a formal spin wave theory for $Cs_2CuAl_4O_8$, which is described by
a Heisenberg alternating spin chain with next nearest neighbor spin-exchange interactions ($e.g.,$ Eq.\ref{eq1}), is no simple and the foremost objective
appears is to find the suitable reference state.
%As mentioned, for such Hamiltonian, an AF or even a doubly staggered AF state does not,
%in general,  stands for a stable reference state. 
For a classical spin system, a proper spin reference state can be found out using a Luttinger-Tisza method\cite{lut-tiz}. 
In our quantum system, we presume of a helical reference state\cite{saptarshi} 
and determine its angle of canting by minimizing the zero point energy.

For our full $J-J'-J_{nnn}$ model, we consider a co-planar helical spin reference state where the spin quantization direction at each site gets tilted by angle $\phi$, say in the spin 
$x-z$ plane,
with respect to its neighboring spin in the chain (see Fig. 1). The spin-coordinates at each lattice point in the helical state is then 
rotated by the canting angle, so that it mimics a ferromagnetic state along the $\hat z$ axis in the
rotated frame\cite{keola}. A successive Holstein-Primakoff transformation in these
new frame gives the Hamiltonian in terms of the spin-wave operators where
the constant zero-energy part of the diagonalized Hamiltonian becomes a function of the canting angle. This is then
 minimized to realize the optimum canting angle to be $\phi=$cos$^{-1}(\frac{2J+J'}{12J_{nnn}})$. 
However, we find that stable magnon modes are obtained in the absence of bond alternations ($i.e.,$ for $J=J'$) alone.
A few results of the magnon modes are shown in Fig.\ref{mag-modes}.
It shows acoustic as well as optical modes.
The small energy gaps in $Cs_2CuAl_4O_8$ as perceived in Quantum Monte Carlo and Exact Diagonalization calculations\cite{main}, however, are not realized in the SWT approximation.
%\vspace {.4 in}

\section{Summary}

As SWT with our reference state chosen gives unstable low energy modes for the bond-alternating chains, 
we need to look for alternative routes to find the excitation in these scenario.Various methods such as 
bosonization\cite{totsuka}, pseudo-boson techniques\cite{cow} or conformal field theory\cite{sakai} has been applied 
to bond alternating problems to show that dimerization of the spin system
develops due to bond alternations and spin-peierls like states emerge with singlet ground state and triplet excitations.
Our model, however, contain many more interactions than it was considered in these works and we wish to to work on similar 
directions and get the actual expressions of the excitation modes for our case in a later communication.

\end{document}